\documentclass[12pt]{article}
\usepackage{graphicx}
\usepackage{amssymb}

\begin{document}

\begin{center}

{\Large
A time-resolution study with a plastic scintillator \\
read out by a Geiger-mode Avalanche Photodiode
}

\vspace{2ex}
A.\,Stoykov, R.\,Scheuermann, K.\,Sedlak

\vspace{2ex}

Paul Scherrer Institut, CH-5232 Villigen PSI, Switzerland \\[1ex]

\end{center}

\vspace{2ex}
\noindent
In this work we attempt to establish the best time resolution
attainable with a scintillation counter consisting of
a plastic scintillator read out by a Geiger-mode Avalanche Photodiode.
The measured time resolution is inversely proportional to the square root of
the energy deposited in the scintillator, and scales to $\sigma = 18$\,ps at 1\,MeV.
This result competes with the best ones reported for photomultiplier tubes.

\vspace{2ex}
\section{Introduction}
High time resolution detection of ionising radiation is an essential prerequisite for
a variety of applications and has been a subject of
numerous theoretical and experimental studies reviewed in \cite{Moszynski79}.
The best time resolution of $\sigma\sqrt{E} = 19\,{\rm ps}\cdot{\rm MeV}^{0.5}$,
where $E$ is the energy deposited in the scintillator,
has been achieved using plastic scintillators read out by photomultiplier tubes (PMTs)
with the signals processed by constant fraction discriminators~\cite{Moszynski79,Moszynski93}.

In this work we focus on achieving the best timing using multipixel
Geiger-mode Avalanche Photodiodes (G-APDs) \cite{Renker_Beaune2005}.
The method is based on the one described in \cite{Moszynski79},
i.e.\ on measuring the time resolution as a function of signal amplitude (deposited energy)
by selecting narrow amplitude windows out of the signal spectrum.
Based on our previous tests done with blue-sensitive G-APDs of different producers,
for the present study we have chosen a Hamamatsu MPPC S10362-33-050,
presumably the best device for fast timing applications currently on the market.

\section{Measurements}
The measurement setup and the scheme of the data acquisition system are shown
in Fig.~1. The setup consists of two identical scintillation counters C1 and C2
detecting electrons from a $^{90}$Sr radioactive source.
In both counters 3\,x\,3\,x\,2\,mm$^3$ BC-422 plastic scintillators are used.
Each scintillator is wrapped in Teflon tape and coupled using optical grease
by a 2\,x\,3\,mm$^2$ face to the 3\,x\,3\,mm$^2$ active area G-APD.
The G-APD is subsequently connected to a broad-band amplifier.
The scheme of the amplifier, its linearity curve, and averaged shapes of the output signals
are shown in Fig.~2.
Both G-APDs are operated at room temperature with a dark current of $I_0 = 1.0\,\mu$A.
The stability of the temperature during the measurements was within 1\,$^{\rm o}$C.

The data acquisition system, depicted in the lower part of Fig.~1,
allowed us to measure amplitude spectra of the analog signals from
C1 and C2, and subsequently to select narrow intervals (windows)
from these spectra, as indicated in Fig.~3.
This way the time difference between
the timing signals of the two counters can be measured
for arbitrary sets of amplitude windows.

First of all, the time resolution of the reference counter C1 was determined for the
amplitude window indicated by the dashed line in Fig.~3.
Both  counters C1 and C2 should have the same time resolution,
since both counters are identical, they are operated under the same conditions, and
identical amplitude windows are selected for both of them.
Fig.~4 shows the measured time-difference spectrum between C1 and C2.
Deduced from this data, the time resolution of the reference counter is 29~ps.

Next, the amplitude window of C1 is kept fixed,
while the amplitude window of C2 scans the amplitude spectrum over the range indicated in Fig.~3.
This way the time resolution $\sigma$ of the counter C2 is measured as a function of
the signal amplitude $A$.
The raw data of this measurement are presented in Fig.~5.

After the measurement of the raw $\sigma(A)$ dependence, the data were corrected
for the non-linearity effects in three subsequent steps:
1)~the measured amplitude $A$ is corrected for the non-linearity of the amplifier
using the equation specified in Fig.~2;
2)~the number of fired cells is calculated as
$N_{\rm cell} = A_{\rm lin} / A_{\rm 1c}$,
where $A_{\rm lin}$ is the linearized amplitude from the previous step, and
$A_{\rm 1c}$ is the charge released at the breakdown of a single G-APD cell (see Fig.~6);
3)~the number of primary photoelectrons $N_{\rm phe}$
is calculated as~\cite{Stoykov_JINST07}:

\begin{equation}
N_{\rm phe} = -\,\frac{m}{\alpha} \ ln\left(1 - \frac{N_{\rm cell}}{m}\right) \ ,
\label{equationNphe}
\end{equation}
\noindent
where $m = 2400$ is the number of G-APD cells per 6\,mm$^2$
(the 2\,x\,3\,mm$^2$ area of the G-APD attached to the scintillator),
and $\alpha = 1.12$ is the mean number of cells fired by a single charge carrier,
as described in Fig.~6.

The obtained number of photoelectrons includes all the non-linearity corrections
(saturation of the amplifier and the G-APD) and
is expected to be proportional to the energy deposited in the scintillator.
The corresponding scaling factor of 0.44\,keV/phe was obtained from the comparison
of the measured amplitude spectrum of C2 with that obtained from GEANT4 \cite{GEANT4} simulations.
In this case the measurement was done in
a ``reverse'' geometry\footnote{Performing the calibration measurement in
the ``reversed'' geometry allowed us to determine the scaling factor for the same counter,
for which the $\sigma(A)$\,-\,dependence was measured.},
i.e.\ with the $^{90}$Sr source positioned in front of the counter C2.
The spectrum was corrected for the non-linearity effects, and then matched with
the spectrum of the deposited energies obtained from GEANT4 simulations,
as illustrated in Fig.~7.  The simulation was done using a $^{90}$Sr source
and the same geometry as in the real experiment. It also indicates that
the variations of the electron time-of-flight are smaller than 4\,ps, and thus
can be neglected in the presented analysis.

The final result for the time resolution versus the number of photoelectrons
and deposited energy is shown in Fig.~8. The energy dependence of the time resolution
follows the form of
$\sigma = \sigma_{\rm 1MeV}/\sqrt{E}$,
with $\sigma_{\rm 1MeV} = 18\,{\rm ps}\cdot{\rm MeV}^{0.5}$.
The dominant contribution to the systematic error comes from the
uncertainty in establishing the correspondence between the measured amplitude
and the deposited energy obtained from the simulations.
A conservative estimate of this uncertainty in the order of 5\,\% leads to the
systematic error in $\sigma_{\rm 1MeV}$ of $\sim 2$\,\%.

\section*{Summary}
Using a small detector made of a fast BC-422 (analog NE-111) plastic
scintillator and a Geiger-mode Avalanche Photodiode we measured a time resolution
of $18\,{\rm ps}\cdot{\rm MeV}^{0.5}$.
This value competes with the best result of $19\,{\rm ps}\cdot{\rm MeV}^{0.5}$ reported for PMTs.
Taking into account compactness of G-APDs and their insensitivity to magnetic fields,
the obtained result promises a breakthrough in fast-timing detection of ionising particles
especially in the presence of high magnetic fields.

The performance of G-APD based detectors in magnetic fields up to 5~T have already been
demonstrated in \cite{Stoykov_PhysB404-HMF},
and recently we were able to prove it also up to 9.5~T.
The details of those measurements will be reported elsewhere.


\clearpage
\newpage
\begin{figure}[htb]
\centering
\includegraphics[width=0.9\columnwidth,clip]{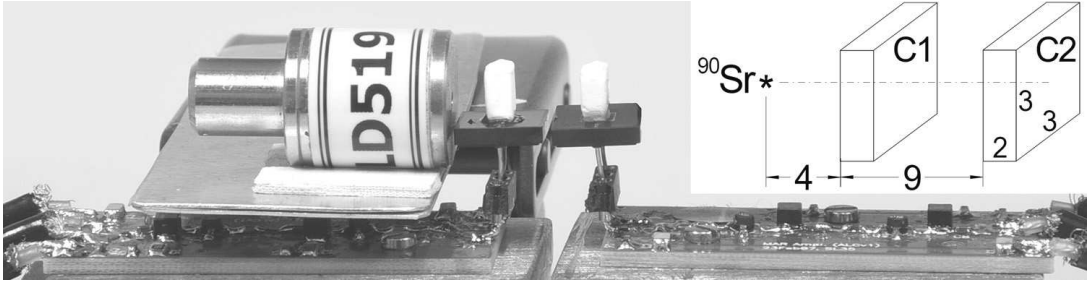}\\[4ex]
\includegraphics[width=0.9\columnwidth,clip]{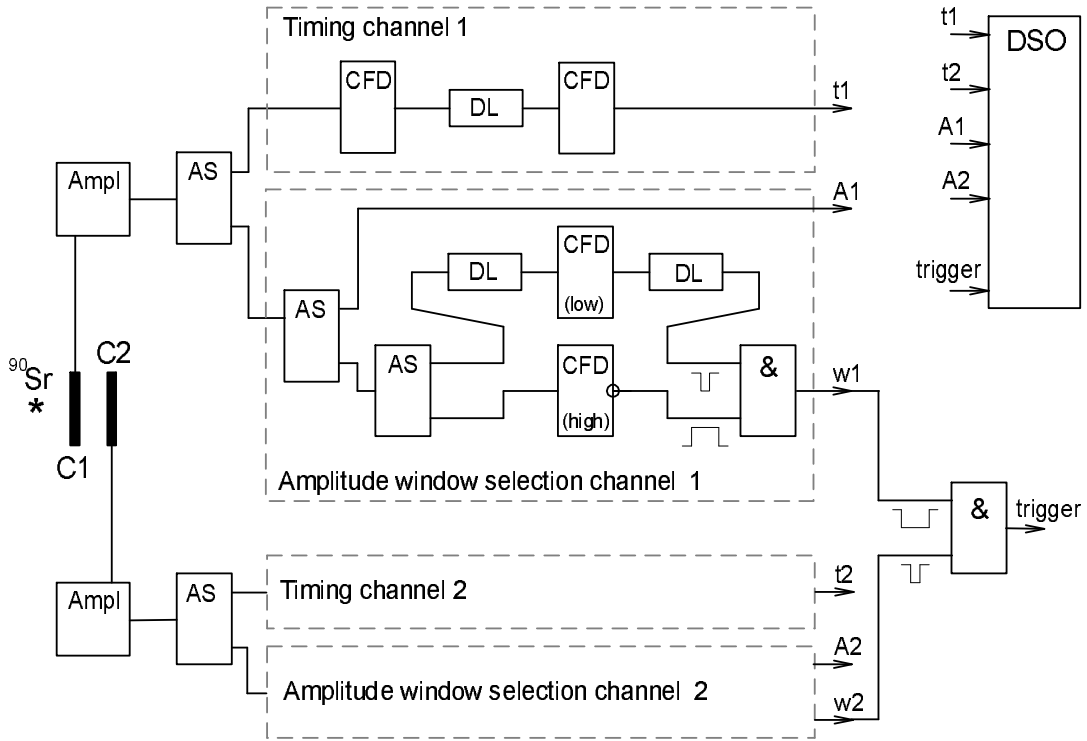}
\caption{(Top)~Measurement setup consisting of
a $^{90}$Sr radioactive source, a reference detector C1, and a detector under test, C2.
The same 3\,x\,3\,x\,2\,mm$^3$ BC422 plastic scintillators are used in both detectors. \
(Bottom)~Scheme of the data acquisition:
AS -- active analog splitter type PSI SP-950 (gain 1, bw 600\,MHz) \cite{PSIVME};
CFD -- constant fraction discriminator type PSI CFD-950 \cite{PSIVME};
DL -- delay line;
\& -- coincidence logic;
DSO -- digital oscilloscope LeCroy WavePro960.
The analog signals of C1 and C2 are split into three paths:
the first signal goes directly to the oscilloscope for the amplitude analysis
(signals A1 and A2);
from the second signal a timing signal is generated (``Timing channels" t1 and t2);
and the third signal, falling into a specified amplitude range
defined by the single channel amplitude analyser (``Amplitude window selection channels"
w1 and w2) generates a trigger signal.
}
\end{figure}

\clearpage
\newpage
\begin{figure}[htb]
\centering
\includegraphics[width=0.8\columnwidth,clip]{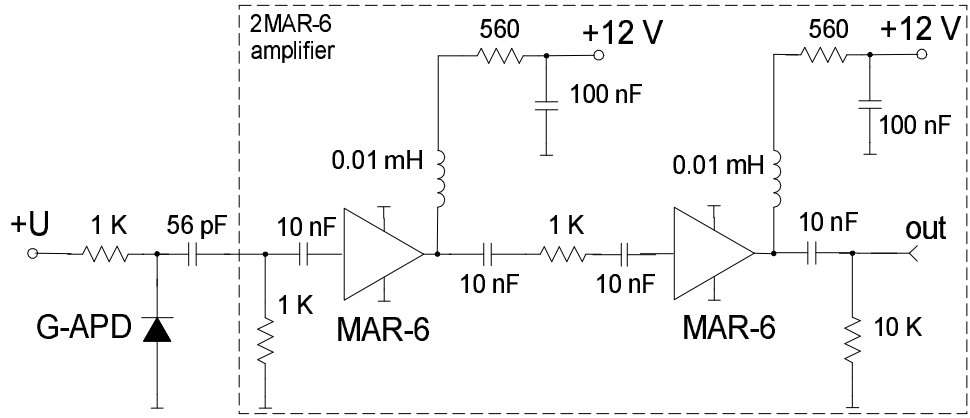}\\[3ex]
\includegraphics[width=0.8\columnwidth,clip]{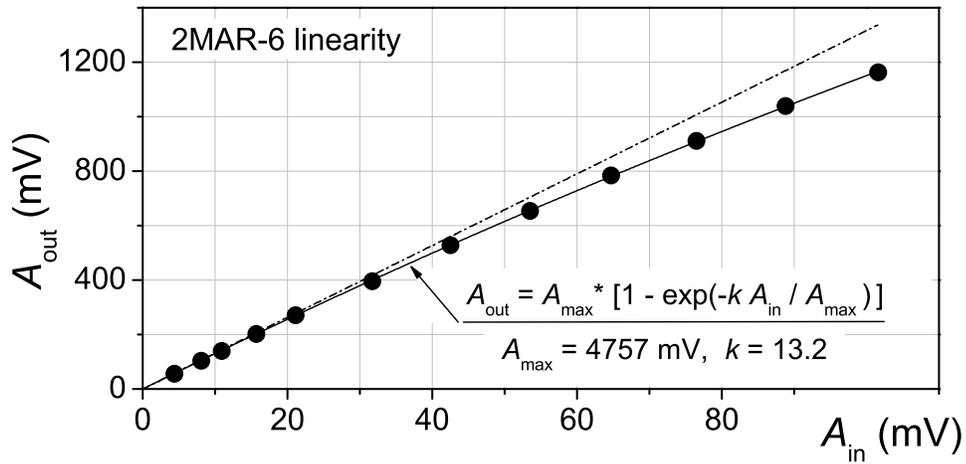}\\[3ex]
\includegraphics[width=0.7\columnwidth,clip]{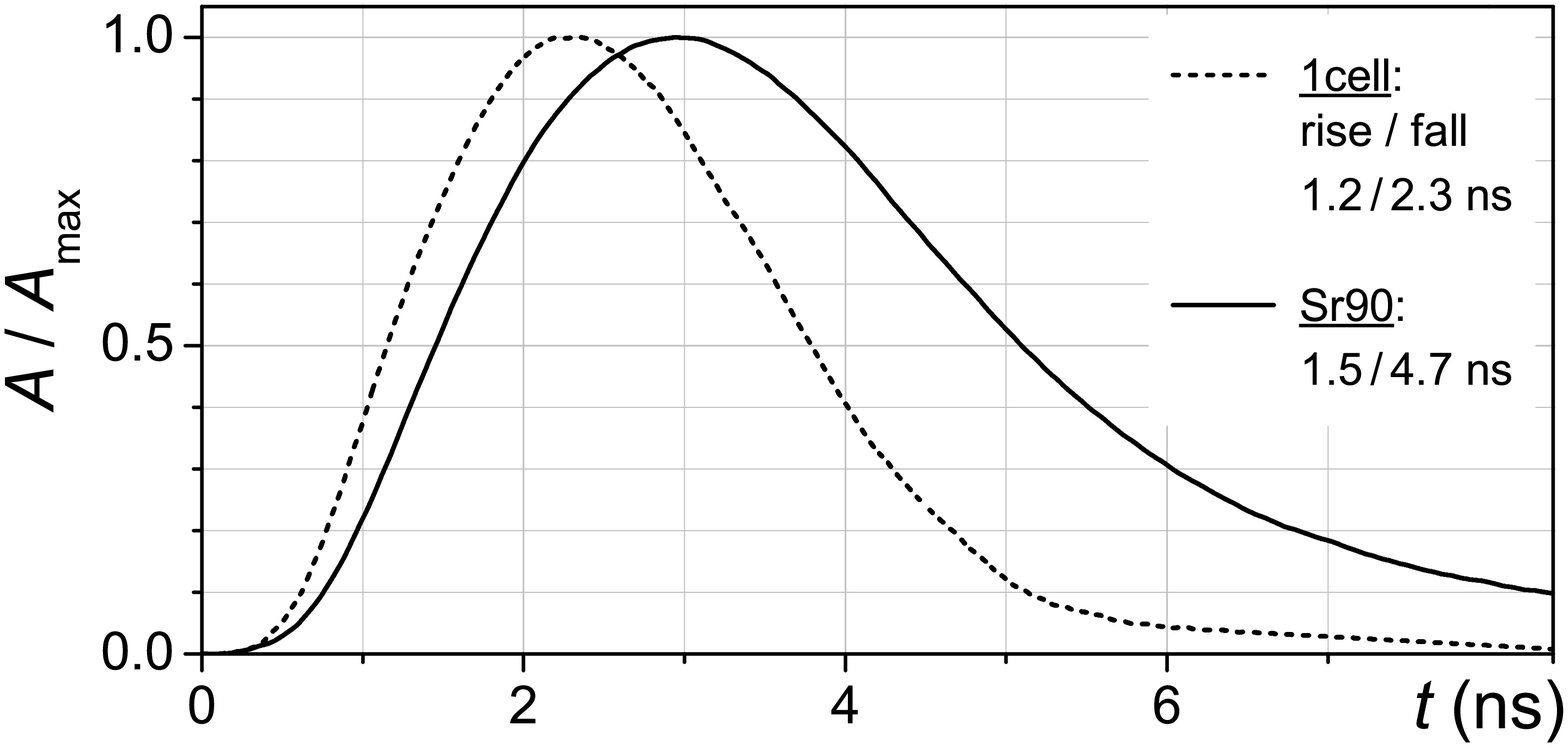}
\caption{(Top)~The biasing scheme for the G-APD and the scheme of the amplifier.
The amplifier is build based on MAR-6 monolithic amplifiers from Mini-Circuits.
(Middle)~Linearity of the amplifier response $A_{\rm out} = f(A_{\rm in})$,
where $A_{\rm out}$ and $A_{\rm in}$  are the pulse heights of input and output signals.
(Bottom)~Averaged shapes of the analog pulses at the amplifier output
from the breakdown of single G-APD cells and from detected scintillations.
The linearity curve was measured with the latter signals.
}
\end{figure}

\clearpage
\newpage
\begin{figure}[t]
\centering
\includegraphics[width=0.85\columnwidth,clip]{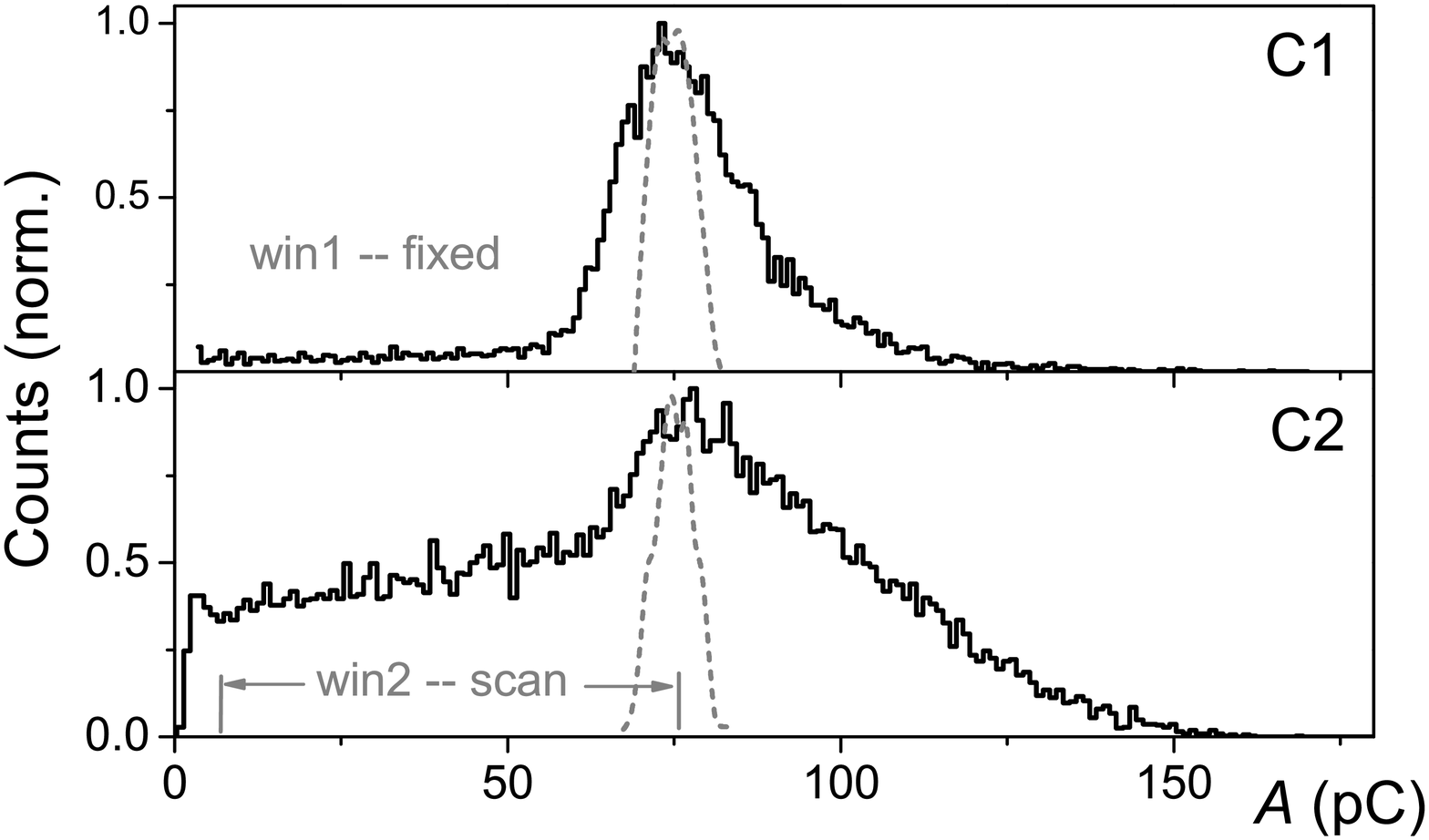}
\caption{Amplitude (charge) spectra of the counters C1 and C2 (solid lines).
The oscilloscope in the both cases is triggered by the signals from C2,
with the C2 window set to ${\rm win2}=(0, \infty)$,
i.e.\ all signals above some low threshold are accepted.
The dashed lines indicate the amplitude windows set in the time resolution measurements:
the window set for the counter C1 (win1) will remain fixed,
while the position of win2 will be varied,
thus allowing us to select different ranges of amplitudes from the spectrum of C2.
The scaling factor from the signal amplitude (charge) to its pulse height
is $\approx 12$~mV/pC.
}
\end{figure}

\begin{figure}[b]
\centering
\includegraphics[width=0.65\columnwidth,clip]{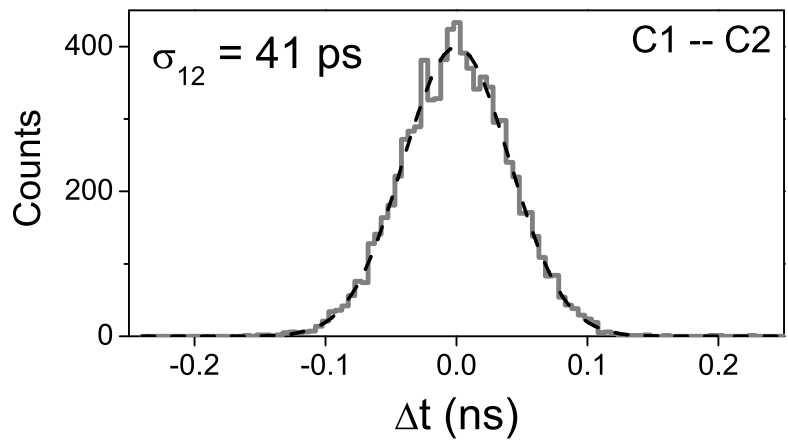}\\[-1ex]
\caption{Time difference between the timing signals of C1 and C2 measured with
the same amplitude windows applied to both counters (i.e.\ win2=win1 in Fig.~3).
The dashed line represents a Gaussian fit to the data.
The time resolution of the reference counter C1 is
$\sigma_1 = 41\,{\rm ps}/\sqrt{2} = 29\,{\rm ps}$.
}
\end{figure}

\clearpage
\newpage
\begin{figure}[t]
\centering
\includegraphics[width=0.85\columnwidth,clip]{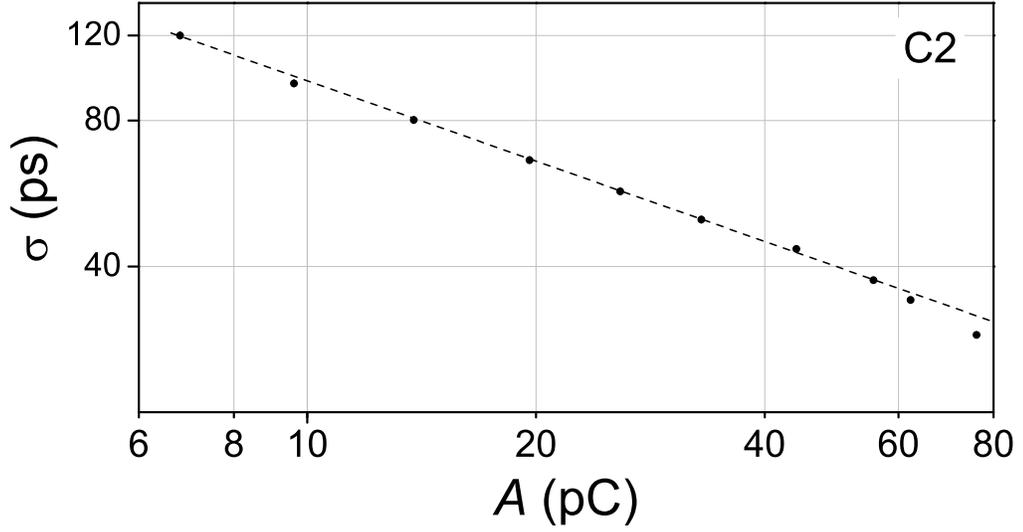}
\caption{Time resolution of the counter C2 as a function of the signal amplitude.
The dashed line serves as a guide for the eye.
}
\end{figure}

\begin{figure}[b]
\centering
\includegraphics[width=0.7\columnwidth,clip]{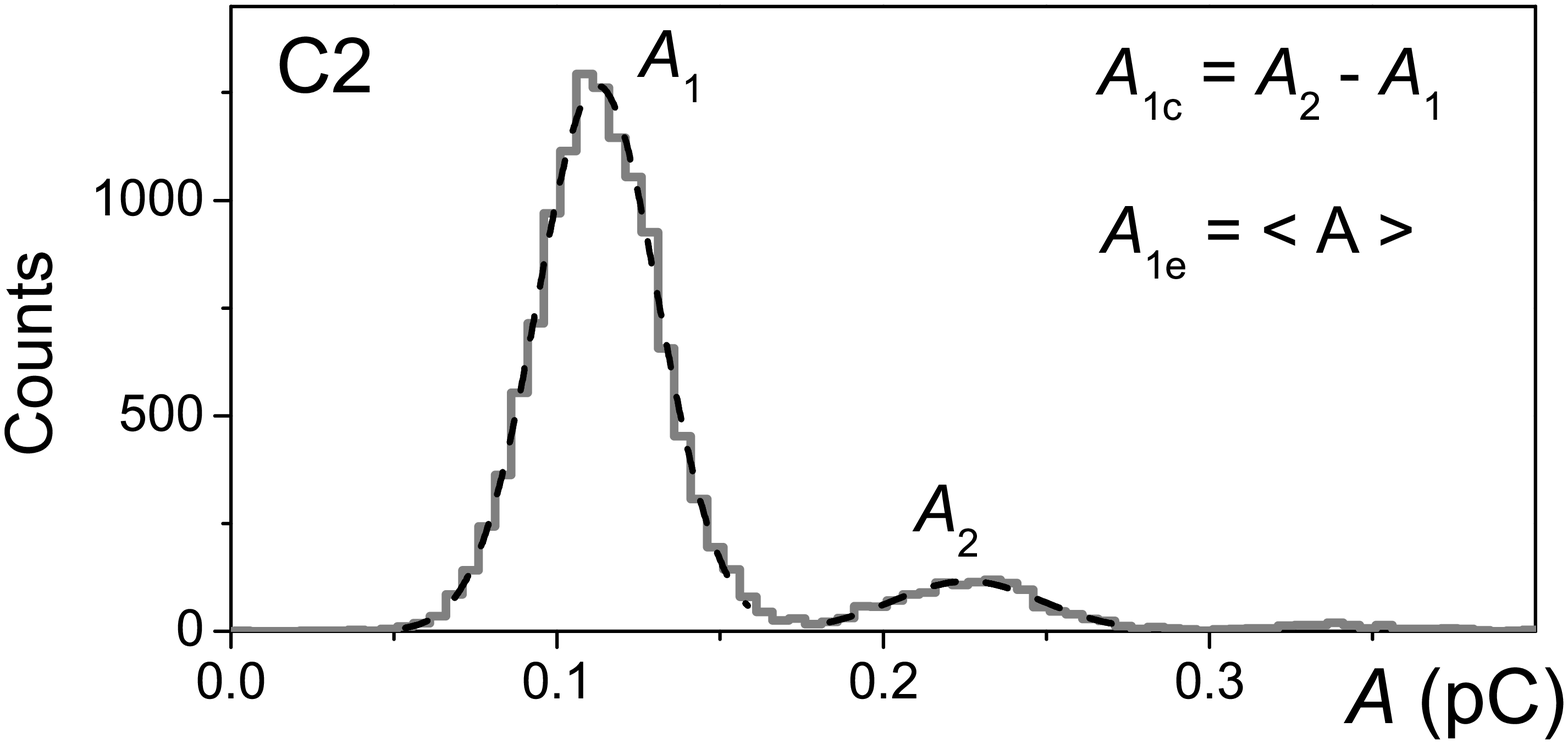}
\caption{Amplitude spectrum of one-electron (1e-) signals of the counter C2.
The 1e-, or ``dark'', signals result from thermal generation of charge carriers
in the active volume of the G-APD. The positions of the peaks corresponding to
one and two simultaneously firing cells are indicated as $A_1$ and $A_2$.
The charge corresponding to one fired cell is calculated as
$A_{\rm 1c} = A_2 - A_1$.
The mean number of cells triggered by a single charge carrier is
$\alpha = A_{\rm 1e}/A_{\rm 1c} = 1.12$, where
$A_{\rm 1e}$ is the mean amplitude calculated over the full spectrum.
}
\end{figure}

\clearpage
\newpage
\begin{figure}[t]
\centering
\includegraphics[width=0.75\columnwidth,clip]{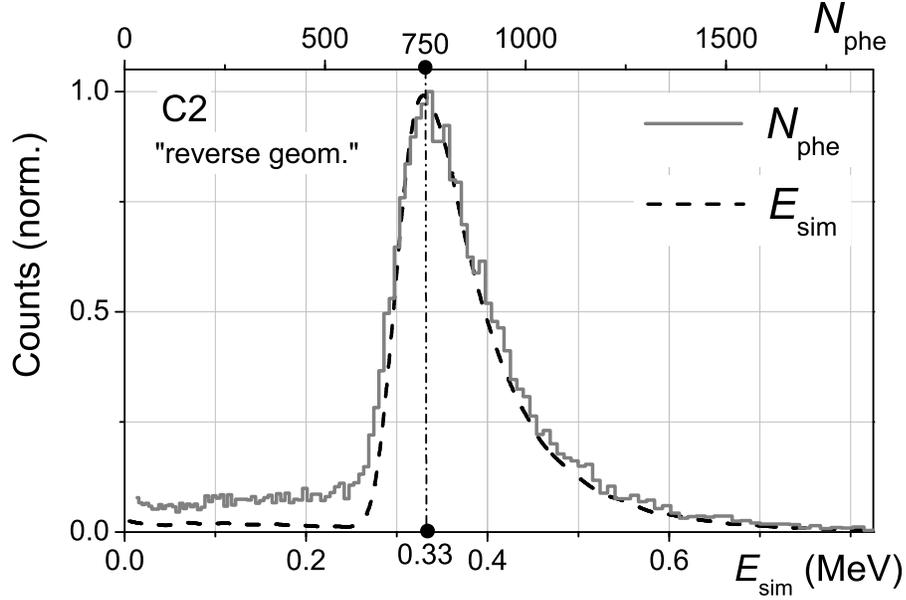}
\caption{Amplitude spectrum of the counter C2 measured in the ``reverse'' geometry
expressed in the number $N_{\rm phe}$ of primary photoelectrons
matched to the spectrum of deposited energies $E_{\rm sim}$
obtained from GEANT4 simulations.
In the reverse geometry, the $^{90}$Sr source is positioned in front of C2 instead of C1,
and the oscilloscope is triggered by C1 with ${\rm win1} = (0, \infty)$.
The most probable signal amplitude of $\approx 750$~photoelectrons corresponds to an
energy loss of $\approx 0.33$\,MeV.
}
\end{figure}

\begin{figure}[b]
\centering
\includegraphics[width=0.85\columnwidth,clip]{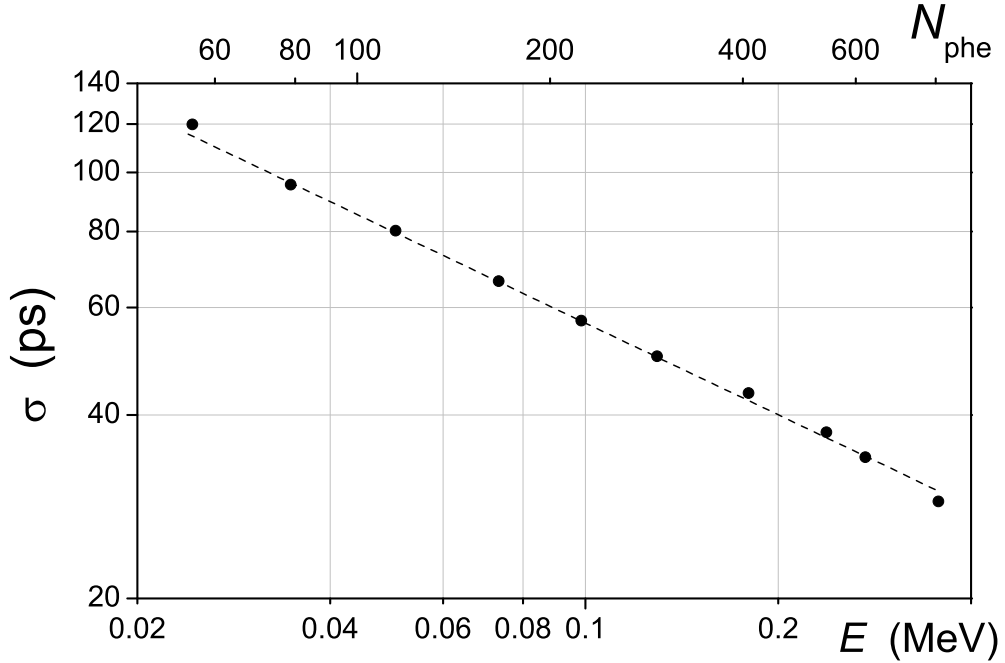}
\caption{Time resolution as a function of energy $E$ deposited in the scintillator
and the number of primary photoelectrons $N_{\rm phe}$
created in the photodetector.
The dashed line is a fit to
$\sigma = \sigma_{\rm 1MeV}/\sqrt{E} = \sigma_{\rm 1phe}/\sqrt{N_{\rm phe}}$,
with $\sigma_{\rm 1MeV} = 17.9(1)\,{\rm ps}\cdot{\rm MeV}^{0.5}$ and
$\sigma_{\rm 1phe} = 850(5)\,{\rm ps}$.
}
\end{figure}

\end{document}